\documentclass{emulateapj}

\usepackage{graphicx}
 \usepackage{natbib}

\def\COBE{{\sl COBE\ }}
\def\um     {\rm \mu m}

\def\etal  {et~al.}
\shortauthors{Fixsen and Kashlinsky} \shorttitle{Cosmic Infrared
Background Dipole} \tighten

\begin{document}

\hfuzz=10pt \overfullrule=0pt

\title{Probing the Universe's Tilt with the Cosmic Infrared Background Dipole}

\author{D.~J.~Fixsen\altaffilmark{1} \& A.~Kashlinsky\altaffilmark{2}}

\altaffiltext{1}{University of Maryland, Goddard Space Flight
Center, MD 20771. dale.j.fixsen@nasa.gov}

\altaffiltext{2}{SSAI, Goddard Space Flight Center, MD 20771.
alexander.kashlinsky@nasa.gov}

\begin{abstract}
Conventional interpretation of the observed cosmic microwave background (CMB) dipole is 
that {\it all} of it is produced by local peculiar motions. Alternative explanations requiring part of 
the dipole to be primordial have received support from measurements of large-scale bulk flows. 
A test of the two hypothesis is whether other cosmic dipoles produced by collapsed structures 
later than last scattering coincide with the CMB dipole. One background is the cosmic infrared 
background (CIB) whose absolute spectrum was measured to $\sim 30\%$ by the \COBE satellite. 
Over the 100 to 500 $\mu$m wavelength range its spectral energy distribution can provide a probe 
of its alignment with CMB. This is tested with the \COBE FIRAS dataset which is available for such 
a measurement because of its low noise and frequency resolution important for Galaxy subtraction. 
Although the FIRAS instrument noise is in principle low enough to determine the CIB dipole, 
the Galactic foreground is sufficiently close spectrally to keep the CIB dipole hidden. 
A similar analysis is performed with DIRBE, which - because of the limited frequency coverage - 
provides a poorer a dataset. We discuss strategies for measuring the CIB dipole with future 
instruments to probe the tilt and apply it to the Planck, Herschel and the proposed Pixie missions. 
We demonstrate that a future FIRAS-like instrument with instrument noise a factor of $\sim10$ lower than FIRAS would make a statistically significant measurement of the CIB dipole. We find that the Planck and Herschel data sets will not allow a robust CIB dipole measurement. 
The Pixie instrument promises a determination of the CIB dipole and its alignment with either the 
CMB dipole or the dipole galaxy acceleration vector. 

\end{abstract}


\section{Introduction}

Due to their origins, the various cosmic backgrounds provide
important information about different aspects of the Universe's
structure and evolution. The adiabatic component of the cosmic
microwave background (CMB), being a leftover from the  Big Bang,
is coupled to the overall structure of
space-time during the last scattering (Turner 1991). On the other
hand, the cosmic infrared background (CIB) and the cosmic X-ray
background (CXB) are produced by emissions from collapsed
structures and trace the later evolution of the universe that took
place at relatively low $z$ (see reviews by Kashlinsky 2005 and
Boldt 1987 for CIB and CXB respectively).

The dipole anisotropy of the CMB is well established from \COBE
FIRAS (Fixsen \etal 1994a) and DMR (Kogut et al 1993, Bennett
\etal\ 1996) measurements. It has a dipole amplitude of $3.346\pm
0.017$~mK in the direction of $(l,b)_{\rm CMB} =  (263.85\pm 0.1,
48.25 \pm 0.04)^\circ$ (Hinshaw \etal\ 2009). If the entire CMB
dipole is kinematic its observed amplitude corresponds to velocity
of $V=370$ km/sec. At least a substantial part of it
must originate from the local motions of the Sun and the Galaxy,
so conventional paradigm has been that {\it all} of the CMB dipole
can be accounted for by motions within the nearby $30-100$~Mpc
neighborhood (see review by Strauss \& Willick 1995). This
paradigm, where CMB dipole converges within the local
neighborhood, has been adopted as standard although several
inconsistencies between various datasets emerged from the
start (Gunn 1988). A possible, if exotic, alternative has been
suggested by Turner (1991, 1992), whereby some of the CMB dipole
is primordial and reflects a tilt across the observable Universe
generated by an isocurvature mode. Such tilt can be produced by
preinflationary remnants pushed very far away by the inflationary
expansion (Turner 1991, 1992; Grischuk 1992; Kashlinsky \etal
1994). In that case the rest-frames of matter and CMB in the
Universe are shifted resulting in the appearance of a net motion
of galaxies with respect to the CMB across the entire cosmological
horizon. Interestingly, this notion has received strong support
from measurements based on the cumulative kinematic
Sunyaev-Zeldovich effect (Kashlinsky  \& Atrio-Barandela 2000)
which indicate a net coherent motion (dubbed the "dark flow") of a
sample of $\sim 1,000$ clusters of galaxies extending to at least
$\sim 500~h^{-1}$Mpc (Kashlinsky et al 2008, 2009, 2010).

If the Universe is tilted, the rest frames of the CMB and galaxies
are shifted and the dipoles of the CMB and CIB/CXB may not
coincide. (There must always be at least a partial
overlap between the dipoles because they share the local motion by
the Sun and the Galaxy). This provides an independent test of the
tilt. The situation with CXB based on HEAO measurements is
inconclusive although the results are marginally consistent with
the CMB dipole (Scharf \etal\ 2000, Boughn \etal\ 2002). The
far-IR CIB presents another opportunity to test this hypothesis.
Aside from testing the alignment of CMB and the dipoles from
diffuse backgrounds originating from Galaxy emission, various
other independent tests of the dark flow phenomenon have been
suggested recently (Itoh et al 2009, Zhang 2010, Kosowsky \&
Kahniashvili 2010).

The far-IR CIB (Puget \etal\ 1996, Schlegel et al 1998, Hauser
\etal\ 1998, Fixsen \etal\ 1998) has been reliably measured, both
its amplitude and its spectral-energy distribution. It is produced
by emission by cold ($T_d\sim 20$~K) dust components in
galaxies and most of it seems to originate at early times, $z\ga
1$ (Devlin \etal\ 2009). Its spectral energy distribution is such
that the dipole component produced by local motion is amplified,
in relative terms, over that of the CMB (see Sec. 3.2.4 of
Kashlinsky 2005). This provides a potential way of isolating the
CIB dipole component from the local motion and probing its
allignment with that of the CMB.

In this paper we present a test for the alignment of the
CMB and CIB dipoles. We do this using the best currently available
data for this type of analysis: 1) the \COBE FIRAS data which have low enough instrument noise and also, importantly, a good frequency coverage with $\Delta \nu\simeq 14$ GHz from 70 to 2,800~GHz across the anticipated peak of the CIB dipole energy distribution; and 2) DIRBE
channel 8-10 datasets which have better angular resolution but limited
frequency coverage and wide band-widths of $\Delta \nu/\nu\sim 0.3-0.4$ at full width. In Sec. 2 we discuss the magnitude of the
CIB dipole assuming perfect alignment with the CMB dipole, our
null hypothesis. Sec. 3 then follows with the FIRAS and DIRBE data
analysis. We demonstrate there that these data come to within a
factor of a few of the null hypothesis CIB dipole, but the
uncertainties of the Galactic modeling prevent more discriminative
determination. In Sec. 5 we discuss details of a hypothetical
experiment that can resolve Galactic contribution better and allow
a more unambiguous measurement. Specifically, we examine the
prospects of the measurement using the Planck,
Herschel and prospective Pixie missions and discuss their potential for measuring the CIB dipole
down to the required accuracy.

\section{Motivation}

The measurements at far-IR ($100~\mu{\rm m} \la \lambda \la
1000~\mu{\rm m}$) using \COBE FIRAS (Puget et al 1996, Fixsen et al
1998) and DIRBE (Hauser \etal\ 1998, Schlegel \etal\ 1998) data
resulted in consistent detections of the CIB. The FIRAS-based
measurements show that the resultant CIB at frequency $\nu$ is
well approximated with (Fixsen \etal\ 1998):
\begin{equation}
 I_\nu^{\rm CIB}=A(\frac{\nu}{\nu_0})^\beta B_\nu(T_d)
\label{eq:cib_sed}
\end{equation}
where $A=(1.3\pm 0.4)\times 10^{-5}, \beta=0.64 \pm 0.12,
T_d=18.5\pm1.2$K, $\nu_0=$ 3 THz and $B_\nu(T)$ is the Planck
function. (The uncertainties in this fit approximation are
correlated.)

The motion at velocity $V$ with respect to the CIB would induce a
CIB dipole of $\delta I_\nu/I_\nu = (3-\alpha_\nu) (V/c)
\cos\theta$ with $\theta$ being the angle to the apex of the
motion and $\alpha_\nu=d\ln I_\nu/d\ln\nu$. Here we ignore the
quadrupole, $O(V^2/c^2)$, and higher contributions resulting from
the relativistic Doppler corrections (Peebles \& Wilkinson 1968).
If the CIB and CMB dipoles are perfectly aligned, the CIB dipole
must lie in the direction of $(l,b)_{\rm CMB}$ and have the
amplitude of:
\begin{equation}
\delta I_\nu^{\rm dipole} = 1.23\times
10^{-3}(3-\alpha_\nu)I_\nu
 \label{eq:cib_dipole}
 \end{equation} Since the dust temperature $T_d \approx 18.5$~K, the CIB does not
reach the Rayleigh-Jeans regime, where the spectral index
$\alpha_\nu\simeq 2$, until $\lambda \ga 400-500~\mu$m. This results
in the significantly negative spectral index of the CIB over much of
the FIRAS and DIRBE probed bands (see. Fig. 1 of
Kashlinsky 2005).

\begin{figure}
\includegraphics[angle=0,trim=30 10 10 10,width=3.3in]{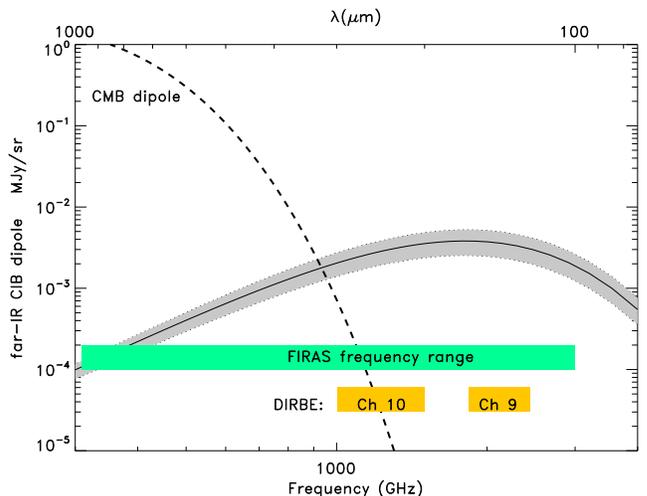}
\caption{Shaded region denotes the uncertainty contours of the spectrum the CIB dipole assuming {\it all} of the CMB dipole is produced by motion within the local volume; its mean value is shown with the black solid line. Dashes show the measured CMB dipole. The continous frequency coverage of FIRAS
and the two lowest frequency (broad) channels of DIRBE are marked.}
\label{fig:cib_dipole}
\end{figure}

Fig. \ref{fig:cib_dipole} shows the predicted CIB dipole spectrum,
using eq. \ref{eq:cib_sed} and assuming perfect alignment with the
CMB; the dipole must have a peak value of $(3-5)\times
10^{-3}$~MJy/sr at 100-300$~\mu$m, or frequencies 1-3 THz. At
longer wavelengths the CMB dipole would overwhelm the signal and
at $\lambda \la 100~\mu$m the CIB dipole decreases, becoming
confused with Galactic and zodiacal light emission. In this
wavelength window, however, because of the slope of its spectral
energy distribution, the dipole in the CIB becomes $\sim10^{-2}$
of its mean level compared to $\sim 10^{-3}$ for the CMB.

We model the CIB and other components as
\begin{equation}
I_\nu(l,b)=I_\nu^{\rm CIB} + \delta I_\nu^{\rm dipole}\cos\beta
+G_\nu(l,b) + {\cal N}_\nu
 \label{eq:CIB_model}
\end{equation}
where $\beta$ is the angle between $(l,b)$ and $(l,b)_{\rm CMB}$,
the mean ($I_\nu^{\rm CIB}$) and dipole ($\delta I_\nu^{\rm
dipole}$) CIB levels are given by eqs.
\ref{eq:cib_sed},\ref{eq:cib_dipole} respectably, $G_\nu$ is the
Galactic emission and ${\cal N}_\nu$ is the noise component at
frequency $\nu$. This expression, coupled with eqs.

\ref{eq:cib_sed},\ref{eq:cib_dipole}, presents our null hypothesis
in the attempt to constrain the CIB dipole, and the tilt, from the
FIRAS data. Note that if all of the CMB
dipole comes from peculiar motions, the CIB dipole amplitude,
$\delta I_\nu^{\rm dipole}$, in this decomposition must be given
by eq. \ref{eq:cib_dipole} and Fig. \ref{fig:cib_dipole}.

Two datasets, FIRAS and DIRBE, are currently available for such analysis and, in addition,
Planck data will be available shortly. In such measurements Galaxy subtraction is critical and
ability to resolve Galatic lines is important in constructing Galaxy templates. The FIRAS instrument
with its continuous fine frequency resolution was the main motivation for this investigation. Broad
band instruments, such as DIRBE (or Planck) are not capable of supplying accurate Galaxy
templates, a point demonstrated earlier by Piat et al (2002) in the context of prospective Planck
data analysis. E.g. C{\sc ii} line at 157$~\um$ provides a critical template when resolved (as in FIRAS which has bandwidth finer than 1\% around that wavelength), but contaminates broad channels such as Channel 9 of DIRBE. In addition it is important to have frequency coverage on both sides of the peak of the predicted energy spectrum of the CIB dipole. This is illustrated in Fig. \ref{fig:cib_dipole} which shows the continuous frequency coverage of FIRAS and the two broad bands of DIRBE at the lowest frequencies.

\section{CIB dipole estimates from FIRAS data.}

The FIRAS data are absolutely calibrated, but the precision of the
calibration is limited by the small amount of calibration data. Using  the FIRAS data in
differential mode reduces the calibration errors and greatly reduces
the systematic errors as well. A dipole (like the CIB dipole) is differential
on the time scale of one orbit.  On this scale the FIRAS calibration is
much better than the absolute calibration. While the
signal to noise is poor at any single frequency, integrated over the
full FIRAS spectrum the CIB dipole is expected to have a signal to
noise ratio of $\sim$3 with 80\% of the statistical weight between
.8 and 1.45 THz. The FIRAS data have poor angular
resolution ($7^\circ$) resulting in a significant Galaxy
contribution at the accessible channels of 100 -- 1000~$\um$.
FIRAS frequency resolution is 13.6~ GHz allowing continuous coverage
between 0.6  and 2.5~THz. The FIRAS instrument resolves the Galactic C{\sc ii}, N{\sc ii}
and CO lines well enough to separate them from the dust continuum enabling the
use of them as templates.

The DIRBE Channels 8,9,10 (nominal wavelengths of 100, 140 and 240~$\um$)
data probe the peak of the CIB spectrum, and have significantly better
angular resolution of $0.7^\circ$, but their filters are broad with $\Delta\nu/\nu \sim
0.3-0.4$ and have substantial levels of noise which is even higher in
Channel 9.

There are several difficulties in extracting the dipole of the
CIB. First the CIB is substantially smaller than the CMB which
results in a significantly fainter dipole.  Perhaps even more
important, the CIB spectrum is much like the Galactic spectrum
since the former includes all dusty galaxies with dust properties
similar to those in the Milky Way. This limits the usefulness of
spectral filters in separating the CIB produced by sources near
$z\sim 0$ from the local Galactic signal. Finally, the
uncertainties of the FIRAS data, both the raw noise and the
systematic errors, rise significantly at higher frequencies.

In this section we present the results of CIB dipole analysis from
the FIRAS and DIRBE maps in Galactic coordinates. In this coordinate
system, $X$ points in the direction of the center of the Galaxy, $Z$
points in the direction of the Galactic North pole and $Y$ is
perpendicular to $X$ and $Z$. Since the solar system is moving faster
in its orbit around the Galaxy (presumably it will be moving slower in
100 million years or so) Y points upwind in the interstellar medium.
The amplitude of this dipole should be that of figure 1 and the direction
should match the direction of the CMB dipole if both dipoles are due to the
motion of the Sun or the dipoles in each direction $(X,Y,Z)$ should match.

\subsection{Templates}

One method of foreground subtraction is the use of
templates. Templates are powerful because the FIRAS full sky maps
have 6,063 pixels so many templates can be used while using
only a small fraction of the number of degrees of freedom
available. Since the cost of adding an extra template is
relatively low, we use many templates.

The key with templates is that they must be attached to the Galaxy.
Using templates that are not attached to the Galaxy runs the risk of
removing the dipole along with the Galactic signal. Next we will
discuss several templates. These templates can be seen in figure 1 of
Fixsen (2009). Except for the zodiacal model all of the templates
look approximately alike.

{\bf i}. One foreground at more than 3~THz is the emission
from the solar system zodiacal dust. We use the zodiacal
model from the DIRBE team (Kelsall \etal\ 1998) as a template for
this emission. 

{\bf ii}. The C{\sc ii} emission from the FIRAS maps (Fixsen \etal\
1999) has several advantages. The data is already matched in beam
shape. Since the C{\sc ii} line from distant galaxies will be
redshifted the C{\sc ii} template is fixed to the local frame (even
if it might excise nearby galaxies as well as the Milky Way).
Kogut \etal\ (2009) note that the square root of the C{\sc ii}
emission tracks emission better than the C{\sc ii} itself. Here both are used.

{\bf iii}. The 408~MHz map (Haslam 1981) is often used as a tracer
of synchrotron radiation. The CIB is dominated by dust with a peak
emission at $\sim 1.5$~THz.  At these frequencies, synchrotron
emission is insignificant. Furthermore there is significant
extragalactic radio emission (Fixsen \etal\ 2010) so there could
be a significant radio dipole. Thus the 408~MHz radio map needs to
be treated with care. Still the North Galactic Spur is clearer in
the radio than with other templates and the Spur is clearly
attached to our own Galaxy. Perhaps there are subtle differences
in the dust associated with this region as well. The
408~MHz map needs to be convolved with the FIRAS beam to be used
as a template. An estimate of the extragalactic
background is subtracted from the 408~MHz map. The major effect of 
the background subtraction is to reduce the coupling to uniform template. As a
uniform background is nearly orthogonal to a dipole.

{\bf iv}. Hydrogen emission, H{\sc i}, should be an ideal tracer of material
in the Galaxy. However in places this line becomes optically thick and so
suffers from self absorption. One way to mitigate this problem is to include
the square of the H{\sc i} as well. The H{\sc i}  from Stark \etal\ (1992) is
convolved with the FIRAS beam to form a template along with map
squared.

{\bf v}. Often the Galaxy is considered as a disk.  In this model
the emission is expected to be distributed as csc$|b|$. Such
modeling has been successfully applied to the DIRBE data isolating
CIB fluctuations in the near-IR, but leading to only upper limits
at wavelengths longer than $\sim 10~\mu$m (Kashlinsky \&
Odenwald 2000). The model has intrinsic difficulties on the
Galactic plane. However, the Galactic plane is too complicated to
realistically model anyway, so the csc$|b|$ template can be used
and the divergence at $b=0$ is ignored.

{\bf vi}. The N{\sc ii} emission from FIRAS and the Al$^{26}$ emission
map (Diehl \etal\ 1995) also must be local, so although they are not
nearly as good tracers of the Galaxy as C{\sc ii}, they are included as templates.

{\bf vii}. The DIRBE team also generated a diffuse stellar map to
model the starlight.  The model is clearly attached to the Galaxy
and one would certainly expect that other emission is related to
starlight, so this model is also included.

The cosmological background is nearly uniform.  To model the
uniform background (both the CMB and the CIB) a simple template
which is 1 everywhere is used. In principle this model should be
orthogonal to the dipole and it is nearly so.  However the weights
for the FIRAS data are not uniform and the various cuts will not
necessarily be symmetric so this term will include the absolute
offset or monopole. The uniform background is also convenient in that it
provides a model of the CIB that can be used to compare to any
observed dipole.

To model a general dipole, we use a set of three orthogonal
dipoles $(X,Y,Z)$ in Galactic coordinates defined above. For
each of these the dipole is modeled as cos~$\phi$ where $\phi$
is the angle between the pixel and the direction of the dipole. Any
other coordinates would work as well but this set is convenient
and there is clearly a hierarchy of expectations of contamination.
That is, the $X$ dipole is most likely to be contaminated as it is
sensitive to the differences between the inner Galaxy and the
outer Galaxy, where there are clearly observed differences in
temperature and composition. The $Y$ dipole is sensitive to the
upwind and downwind differences in Galactic radiation.  The $Z$
dipole is sensitive to difference between the north and south
hemispheres of the Galaxy. There is no intrinsic reason for an
asymmetry in this direction although the sun is a bit north of the
Galactic equator. Detailed discussion of the relative weight of
the errors for various cuts and configurations is given by
Atrio-Barandela et al (2010, see  e.g. Fig. 4c there).

In other papers the DIRBE band 8, 9, and 10 are used as templates
 (Fixsen \etal\ 1996, Fixsen \etal\ 1997).
These bands indeed model the dust very well, however, these data
also include the CIB dipole and using them likely would subtract
the dipole along with the local dust emission. Still any one of
these is a sensitive measure of the likely Galactic contamination.
The DIRBE band 10 is used to excise, without bias, the regions of
high Galactic emission.

\subsection{FIRAS fitting}
 Given a collection of template maps, $T$, and the FIRAS data, $D$,
 along with the pixel weights formed into a diagonal matrix, $W$, a
 least squares fit is made separately at each frequency, resulting in
a spectrum, $S$, for each template:
\begin{equation}
S=(T\cdot W \cdot T^t)^{-1} \cdot T \cdot W \cdot D
\label{eq:sub}
\end{equation}
No new correlations are introduced in this process but the frequencies
are mildly correlated by the previous FIRAS calibration process. Fixsen et al (1999) discuss in detail the line templates used here and their spatial properties.


Since most of the weight for the CIB dipole is in the higher frequency data
it is appropriate to use the high frequency weights. All of the Galactic
templates have a similar form, so they are highly correlated. This is
not a severe problem as long as the matrix is not singular and one is
not interested in the spectra associated with the different Galactic
templates.

Even with all of these templates the Galactic plane is far too complicated to
model in detail.  As usual, a cut is made on either the Galactic
latitude (ie data is ignored for $|b|<$~cut), or  on some measure
of Galactic radiation. Here we base our excision on the DIRBE band 10 level.
Ideally, one could change the level of cut and find a level where the
CIB dipole estimate was insensitive to the level of the cut. In fact for the
CMB dipole the direction and amplitude of the fit hardly change going
from cutting 10\% of the data to cutting 50\%.

So we have made fits ranging from no cuts to excising 50\% of the data in 5\% increments.
Cutting beyond 50\% of the data runs into three problems. One is that the support of
the $X$ and $Y$ dipoles is greatly reduced.  A second is that the
already highly correlated Galactic models approach singularity as the most emisive
and distinctive parts of the Galaxy are systematically excised. Finally the
already limited signal to noise is reduced.

Fits can be made with or without each of the 10 foreground templates at
each of 11 different Galactic cut levels leading to $11 \times 2^{10}=11264$
different fits each with several spectra leading to more than $10^5$ spectra.
Rather than attempt to present all of these we will attempt to show the
best fit.

Clearly fits made including the Galactic plane are still contaminated with
foreground emission. Using the CMB dipole as a guide, it appears that
excising 30\% of the data is approximately the optimum place to cut.

With 70\% of the darkest part of the sky, the order of importance of a
single template is: 1) the zodiacal template,  2) the $\sqrt{\rm C{\sc II}}$ map,
3) the 408 MHz map,  4) the  H{\sc i} map, 5) the  csc model, 6) the  C{\sc ii} map,
7) the  N{\sc ii} map,  8) the  Al$^{26}$ map,  9) the  H{\sc i}$^2$ map and
finally 10) the DIRBE stellar map.

The spectra related to each of the templates can appear peculiar to those
unfamiliar with the data and the Galaxy. For example, when including
a C{\sc ii} and an N{\sc ii} template the corresponding
C{\sc ii} and an N{\sc ii} spectra contain their corresponding lines
as one would expect. The C{\sc ii} spectrum vaguely resembles the
mean Galactic spectrum, but the N{\sc ii} spectrum is negative at
low frequencies and positive at high frequencies. This is because
the N{\sc ii} line is more concentrated near the center of the Galaxy
where the average starlight and hence dust temperatures are
higher. The negative part of the N{\sc ii} spectrum is always canceled by
the dominate positive C{\sc ii} spectrum, and the fit uses the
N{\sc ii} to effectively adjust the temperature of the dust.

Since all of the Galactic templates are highly correlated, emission
can easily slosh from one to another depending on whether or
not other templates are present or whether or not some regions of the
sky are included. The high correlation is expected because all
of the material, hence the line emission, is concentrated in the
Galactic plane. This should help with resolving extragalactic
features as most of the high latitude emission is thus local and
hence one might expect it to be quite uniform in form if not
in intensity.

In this fitting we have not modeled errors in the templates although clearly
none of the templates is perfect. Template errors can amplify the ``sloshing''
that is already present. Small scale errors are likely to just add noise, however
large scale errors such as a mismatch in calibration between the north
and south calibration of the H{\sc i} could dominate the error budget for the
$Z$ dipole of the CIB.

Even after subtracting the best fit template spectra there is still significant
residual signal. This often appears as a broad dip around 1 THz with a
corresponding peak about 2 THz in the residual of fits of the uniform spectrum.
This could either be a FIRAS calibration artifact (the absolute spectrum is
not calibrated as well as the differential spectrum) or variations in the dust
spectrum that are not modeled with the selection of templates used here.
The trough and peak are only a sigma or so in amplitude but since this
"feature'' extends over many frequency bins it is significant.

\begin{figure}
\includegraphics[angle=90,trim=10 20 20 100,width=3.5in]{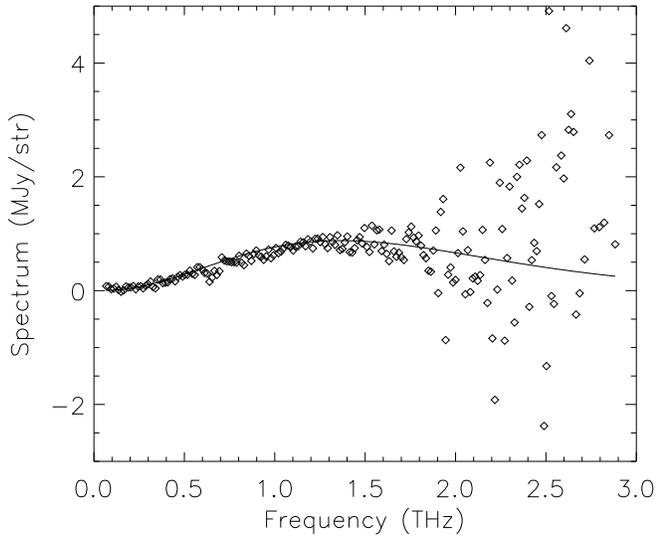}
\caption{The spectrum associated with the uniform template (the background)
with the CMB subtracted. The smooth line
is the CIB spectrum from Fixsen \etal\ 1998. The scatter at high
frequencies are from the uncertainty of the FIRAS data.}
\label{CIB}
\end{figure}

The smooth CMB dominates the uniform spectrum although a
distinct CIB spectrum can be seen at higher frequencies. In fig \ref {CIB},
a single blackbody spectrum has been subtracted from the
uniform background spectrum. This is a good model for the CIB.
The line is  the CIB spectrum from Fixsen \etal\ 1998, which is a reasonable
representation of the data.

\begin{figure}
\includegraphics[angle=0,trim=50 10 10 5,width=3.4in]{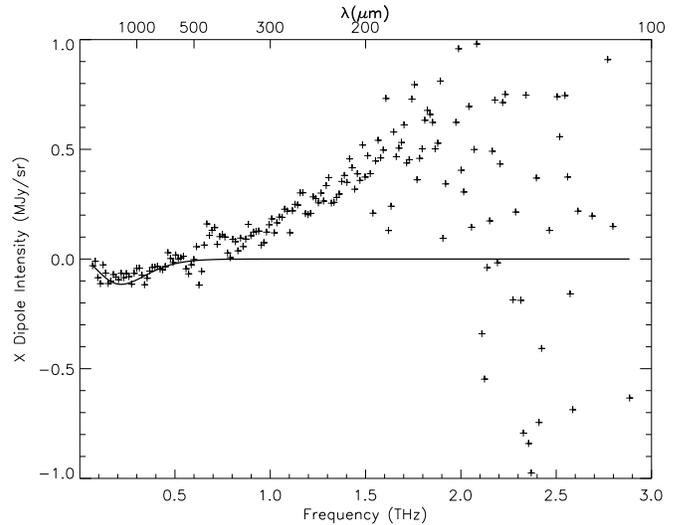}
\caption{The spectrum associated with the X dipole template. Solid line is
the model of the CMB dipole spectrum.} \label{Xpol}
\end{figure}

\begin{figure}
\includegraphics[angle=0,trim=50 10 10 5,width=3.4in]{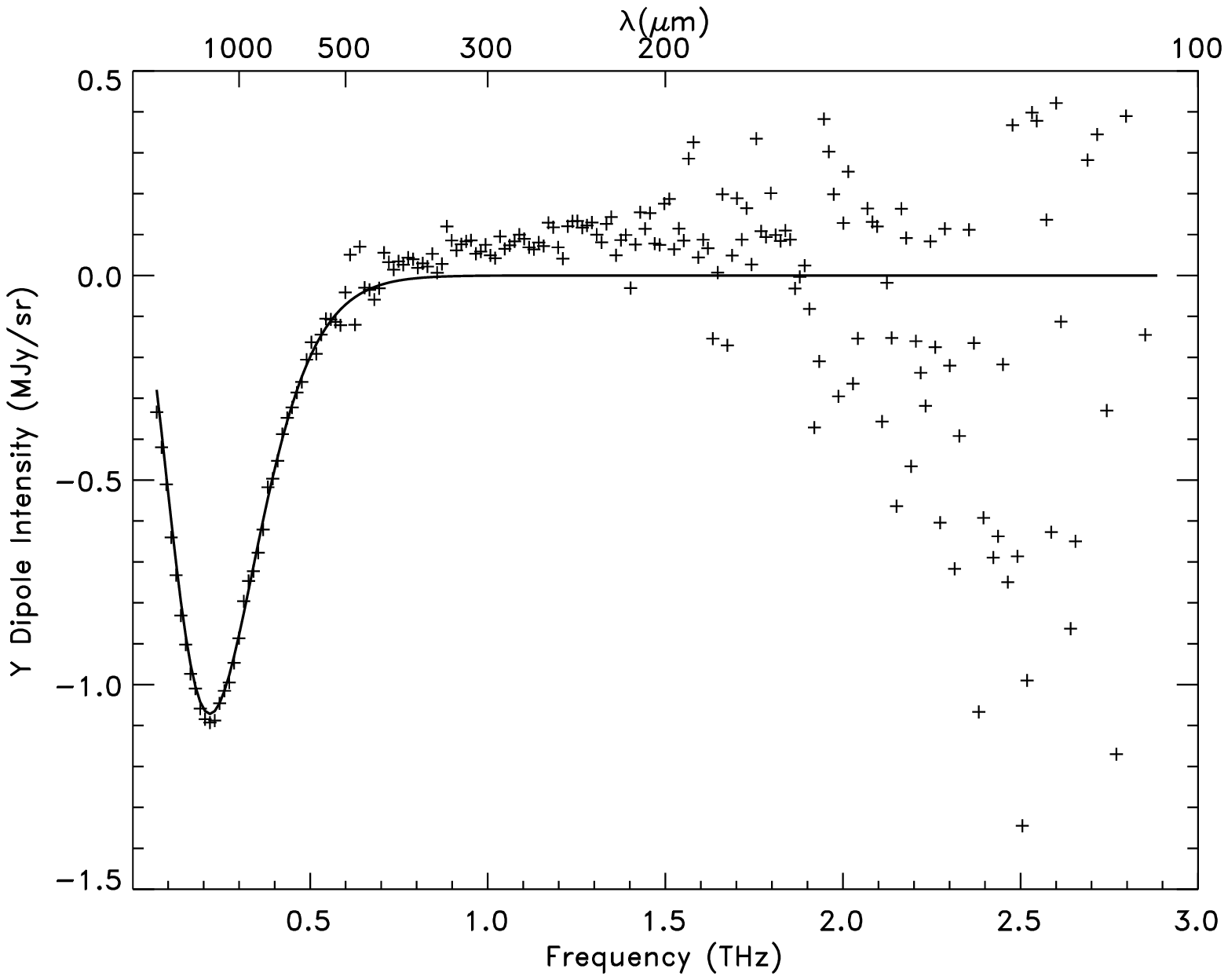}
\caption{The spectrum associated with the Y dipole template. Solid line is
the model of the CMB dipole spectrum.} \label{Ypol}
\end{figure}

\begin{figure}
\includegraphics[angle=0,trim=50 10 10 5,width=3.4in]{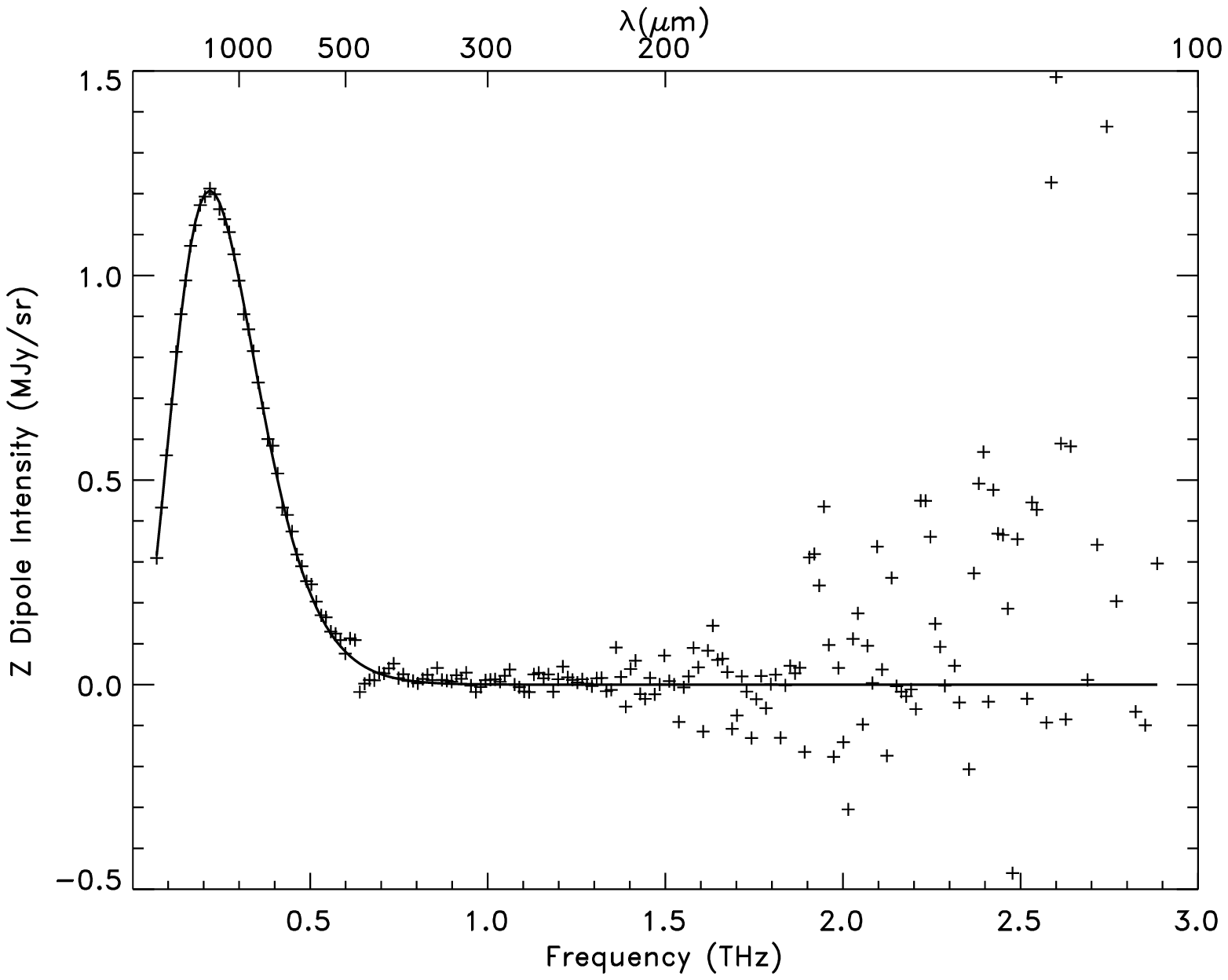}
\caption{The spectrum associated with the Z dipole template. Solid line is
the model of the CMB dipole spectrum.} \label{Zpol}
\end{figure}

Figures \ref{Xpol},\ref{Ypol},\ref{Zpol} show the 3 derived dipole
spectra. The dipole is predicted from the CIB spectrum in figure
\ref {CIB} and the velocity relative to the CMB, $v=370$~km/s
toward $(l,b)_{\rm CMB}$, shown as solid lines.

\begin{figure}
\includegraphics[angle=0,trim=50 10 10 5,width=3.4in]{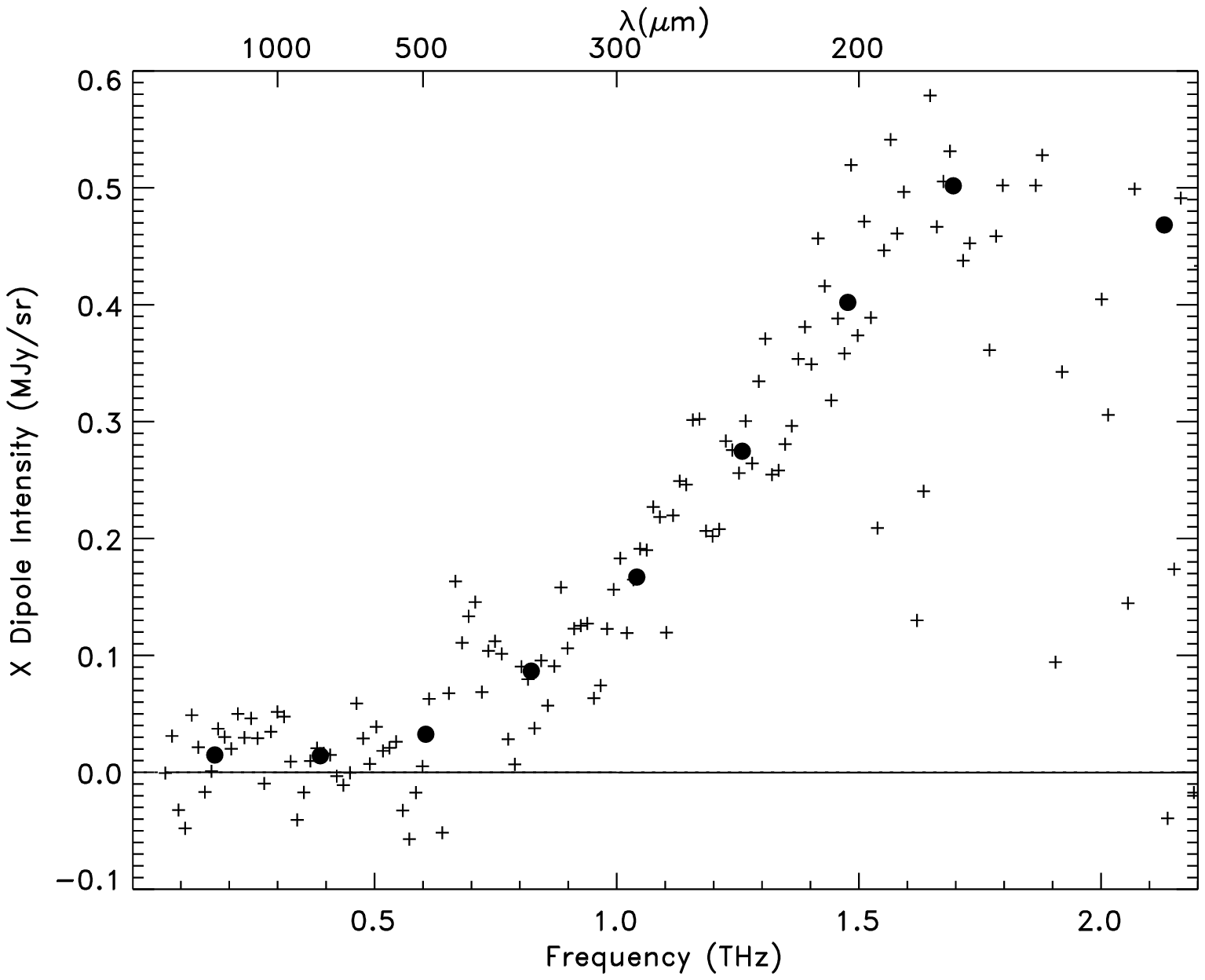}
\caption{The X dipole spectrum with the X CMB dipole spectrum
subtracted. Filled circles show the data with neighboring
frequencies averaged to reduce the noise.
The smooth line is the expected X CIB dipole.} \label{XCIB}
\end{figure}

\begin{figure}
\includegraphics[angle=0,trim=50 10 10 5,width=3.4in]{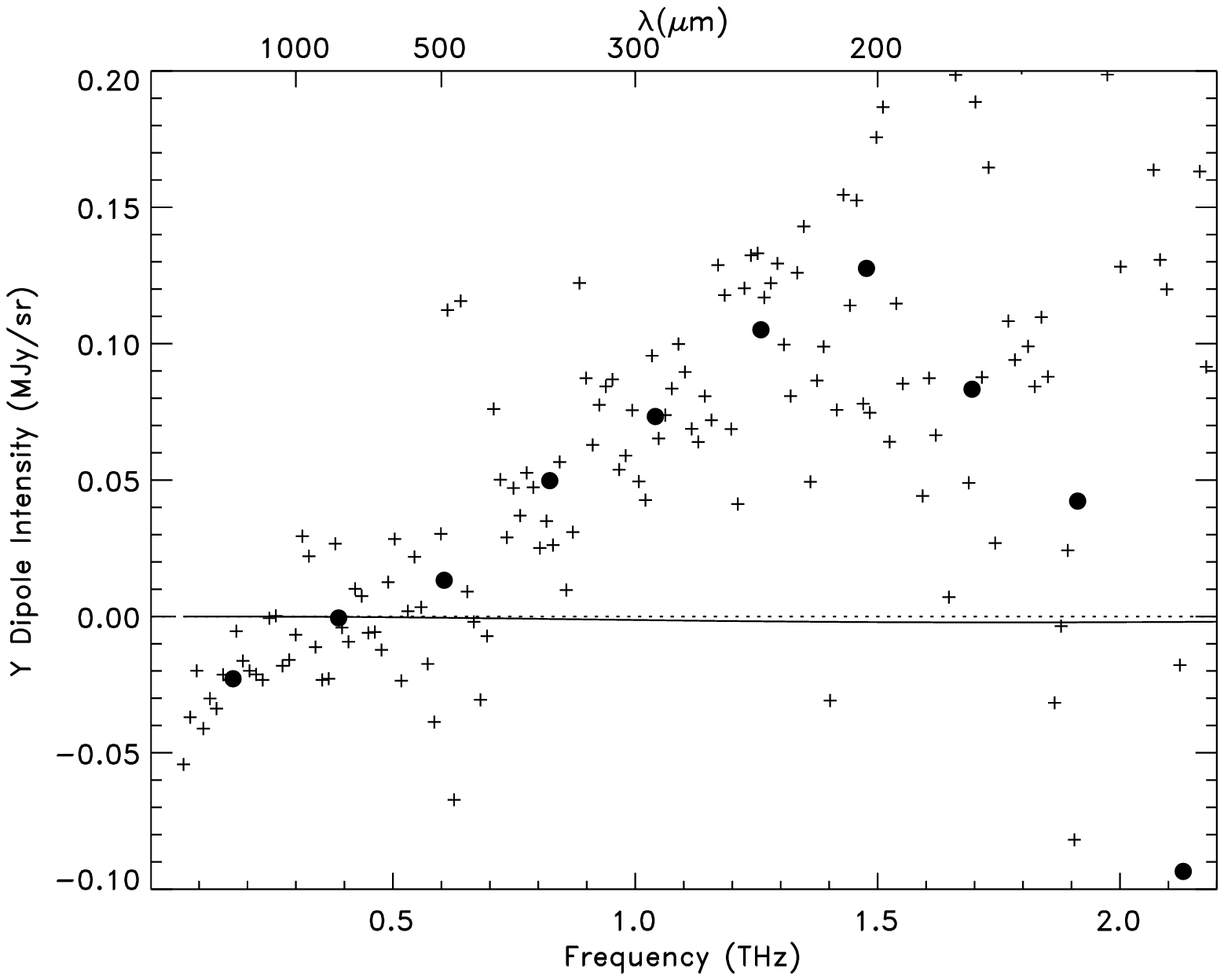}
\caption{The Y dipole spectrum with the Y CMB dipole spectrum
subtracted. Filled circles show the data with neighboring
frequencies averaged to reduce the noise. Note the scale is
different from the X-dipole. The smooth line is the expected Y CIB dipole.} \label{YCIB}
\end{figure}

\begin{figure}
\includegraphics[angle=0,trim=50 10 10 5,width=3.4in]{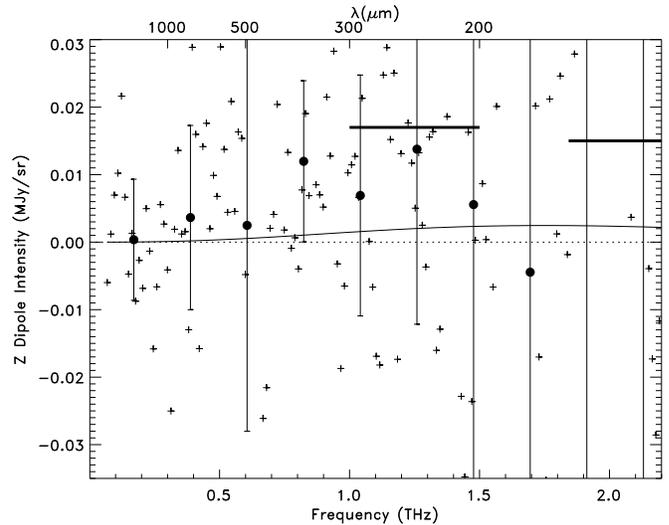}
\caption{The Z dipole spectrum with the Z CMB dipole spectrum
subtracted. Filled circles show the data with neighboring
frequencies averaged to reduce the noise. Note both the horizontal
and vertical scales are different from the previous plots. The line is
the predicted Z CIB dipole spectrum. Three points of the FIRAS data are
above the scale:at 1.92 THz with a dipole value of .36 MJy/sr.
DIRBE data Channels 9 and 10 are shown with horizontal bars of width
corresponding to the respective band filters.} \label{ZCIB}
\end{figure}

The nominal fit to a CIB dipole suggests a dipole several times larger
than would be expected from the CMB dipole. However the fit is not
stable under different cuts and with different templates.  So with
the FIRAS data we conclude the equivalent velocity of the CIB
dipole is less than $7500$~km/s. {\em The
uncertainty is completely dominated by systematic errors in
subtracting the foreground Galaxy emission.}

\section{DIRBE}

The DIRBE data has higher resolution so the number of potential
templates is much larger.  However many of the templates (C{\sc ii}, N{\sc ii}
etc) do not have sufficient resolution to take advantage of the
higher resolution of the DIRBE instrument.

Although the DIRBE instrument has 10 bands (frequencies), the
mid-range and high frequencies have substantial contamination from
the zodiacal dust. In the DIRBE data the mean level of the CIB was
detected in just two bands (bands 10 and 9 or 1.25 and 2.1~THz, Hauser et al
1998).

The 100~$\um$\ (3~THz) band has a nonlinear response. While the effect has
largely been corrected any dipole would be suspect.  The peak
of the CIB dipole should be about 2~THz (150~$\um$), so this band is in the
Wein part of the spectrum. Also there is still significant residual zodiacal
 contamination of this band. Thus this is not a good
channel to look for the CIB dipole.

The 140~$\um$\ (2.1~THz ) and 240~$\um$\ (1.25~THz) bands have many of the same problems
as the FIRAS data (not stable under different cuts, high noise) and these lead to
two numbers rather than the spectrum of FIRAS. The results of the two channels
are respectively 5 and 6 times the expected dipole in the Z direction closely matching
the FIRAS results.  This shows the agreement between the FIRAS and DIRBE
data but the uncertainties  are dominated by the same systematics of the
foreground subtraction that plague the FIRAS result. Since the same templates
were used the systematics are correlated with those of the FIRAS results.

\section{Future CIB Dipole Measurements - experimental parameters and prospects}

As shown in the preceding sections, the CIB dipole must peak at
$\sim200~\um$\ and must be $\sim 5\times 10^{-3}$~MJy/sr if aligned with the CMB dipole.
Although, in principle, the FIRAS noise is low enough to tease out the CIB dipole,
in practice, the main impediment
to probing the alignment of the CIB and CMB dipoles is confusion
with Galactic emission. The physical reason for this is the
fact that - with the FIRAS noise levels and without removal of low-$z$ sources - the bulk of the
CIB comes from galaxies whose spectral energy distributions are
similar to that of the Milky Way. The available templates do not have
sufficient fidelity to remove the contaminating foreground signal. With the
FIRAS and DIRBE datasets one cannot overcome the Galactic
confusion and reliably measure/constrain the CIB dipole, but as we discuss in this section
it is possible to successfully make the measurement with the next generation space missions.

\begin{figure}
\includegraphics[trim=20 10 10 5,width=3.5in]{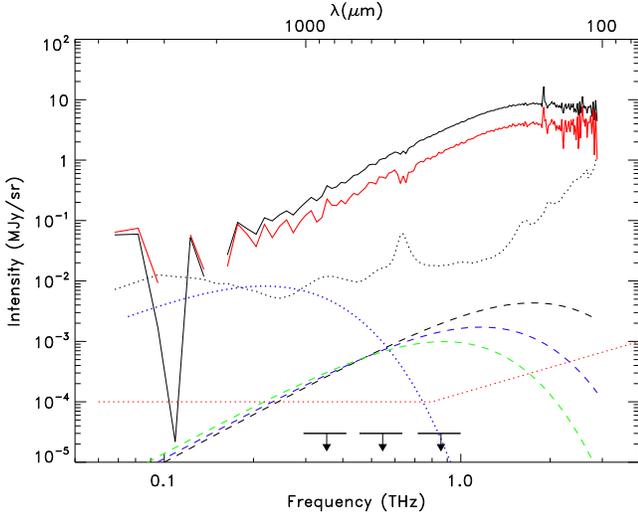}
\caption{The solid black line is the average Galaxy spectrum
for $|b|>10^\circ$  corresponding to the G10 template in the main text. The solid red line is 
the average Galactic spectrum at $|b|>30^\circ$ corresponding to the G30 template in the text. 
The dashed lines are the expected spectra of the CIB dipole; black, blue, green line
correspond to sources at $z\ga 0, 0.5, 1$ respectively. The dotted blue
line shows the CMB dipole residual uncertainty (at 0.017~mK rms
uncertainty). The FIRAS
noise is shown with the dotted black line. An estimate of
the Pixie noise is shown with dotted red line. Three horizontal
bars with arrows show the Planck noise limits (below the plot) at
its three highest frequency channels. } \label{cibde}
\end{figure}

Fig. \ref{cibde} shows the average Galaxy emission compared to
the expected CIB dipole spectrum. The residual CMB dipole, assuming the current uncertainties,
is shown with a dotted blue line and dominates emission beyond $\sim500~\um$.
The displayed CIB dipole is shown assuming contributions from
all galaxies (black) and from galaxies remaining at $z>0.5,1$
(dashed blue and green lines respectively). For
simplicity it was assumed that the CIB is given by eq. 1 and that the apparent dust temperature of the emitters scales as $T_d(0)/(1+z)$ and with sources at $z>0.5,1$ contributing 70,
50\% of the far-IR CIB at 0.6 THz (500~$\um$). The latter normalization is consistent with
the BLAST results in the far-IR which indicate that most ($\ga
70\%$) of the CIB at these wavelengths originates in sources at
$z\ga 1.2$ (Devlin et al 2009). Thus we can
conservatively assume that if we eliminate galaxies to $z=1$ the
remaining CIB at 100-300 $\um$\ would be 50\% of the total. Fig. \ref{cibde}
shows that CIB produced by high-$z$ sources becomes progressively more distinguishable in spectrum at increasing $z$. This still
leaves the CIB dipole significantly below the foreground Galactic
radiation but the spectral difference allows discrimination between
the bluer Galactic spectrum and the redder CIB spectrum.

The currently flying and proposed space missions relevant to the proposed measurement are:
\begin{itemize}
\item {\bf Planck}/HFI has two channels at 545 and 857 GHz (550 and 350~$\um$).
By 2012, {\bf Planck} will have mapped the full sky with $\theta_{\rm
 Planck} = 5^\prime$ resolution. Fig. \ref{cibde} marks the highest frequency Planck channels. In principle, the noise levels should allow detection of the CIB dipole. However with only two frequencies  and the broad channels with $\Delta\nu/\nu=0.33$ for the HFI instrument (Tauber 2004),
convincing separation of the Galactic foreground will be
difficult, a point already realized by Piat \etal\ (2002).
\item {\bf Herschel}/SPIRE covers 250, 350 and 500~$\um$\ (1200, 857 and
545 GHz) with resolution of order 1 arc minute and FOV $4^\prime
\times 8^\prime$ (Griffin \etal\ 2008).  The detectors are the same spider web detectors
as used in the {\bf Planck} instrument
As with {\bf Planck} a few
months of data is enough to get to the noise levels required to
look for the CIB dipole. Observing a few widely spaced regions is
sufficient, in principle, to detect the CIB dipole. But low
frequency noise in the system is likely to limit the calibration
stability of the background levels over widely spaced (in time as
well as space) observations.
\item {\bf Pixie} is a proposed space mission to observe the full sky
with a large throughput (4 cm$^2$ sr) Fourier Transform
Spectrometer (Kogut \etal\ 2010). With full sky coverage at angular resolution of $2.5^\circ$  and 15~GHz frequency resolution, {\bf Pixie} will be able to resolve
spectral features like the 157~$\um$\ C{\sc ii} line to separate the
Galactic foreground from the CIB.  With a noise floor below
$10^{-4}$ MJy/sr per multipole per frequency bin the noise levels are very low to allow a detailed search for CIB
anisotropy. The spectral coverage will also allow improvement
in estimation of the microwave background dipole.
With spectral coverage to 6~THz {\bf Pixie} will be
able to generate new maps for N{\sc ii}  at 122 and 205~$\mu$m,
and O{\sc i} at 63 $\mu$m as well as generate improved C{\sc ii}
and N{\sc ii} maps. Fig. \ref{cibde} shows the estimated noise levels over the frequencies covered by {\bf Pixie}; they are over two orders of magnitude better than FIRAS.
\end{itemize}

The contributions to the measured sky dipole from diffuse maps that can confuse the CIB dipole measurement will be from
\begin{enumerate}
\item {\it Instrument noise}. The dotted lines in Fig. \ref{cibde} show the FIRAS and {\bf Pixie} noise
levels. Summing over all the FIRAS bands give us S/N$\sim3$ if the Galaxy were
eliminated accurately. The uncertainty in the Galaxy templates costs most of this $S/N$
giving only upper limits on the CIB dipole levels.
\item {\it CMB dipole residual contribution}. The CMB dipole is measured with
S/N$\ga 1.5\times 10^4$ down to the residual uncertainty of $\sigma_{\rm CMB}\simeq 0.017$~mK.
The spectrum of this contribution will be given by $B^\prime_\nu = dB_\nu(T_{\rm cmb})/dT_{\rm
CMB}$. The dotted blue line in Fig. \ref{cibde} shows the contribution from the residual CMB dipole
with the current measurement uncertainty, but it can be improved with improved measurement.
\item {\it Galactic contribution} is described by the available templates and is by far the largest
obstacle to the current CIB measurements. This solid black and red lines show the Galactic
foregrounds at Galactic cuts $b_{\rm cut} =10^\circ, 30^\circ$ respectively; we refer to their
templates as G10, G30.
\end{enumerate}

The key requirement is to break the degeneracy between the far-IR
CIB energy spectrum and that of the Galaxy over the wavelengths
where CIB dipole is near its peak. CMB dipole dominates the
long-wavelength emissions, but its energy spectrum is very
accurately known and so it can be subtracted making the residual
small at wavelengths below $\sim 500~\um$. If we were able to
resolve galaxies out to sufficiently high $z$ and remove them from
the maps, the spectrum of the remaining CIB would potentially be
sufficiently different to allow robust removal of the Galactic
contribution to the dipole. Such experiment should be finely
tuned, since at the same time one would need to leave enough
sources in the confusion to generate sufficiently measurable
levels of the far-IR CIB. Or alternatively, the low-$z$ part of
the CIB can be removed together with the Galactic foreground, but
that too requires sufficient resolution to remove the
Galaxy accurately enough.

Given spectal templates of the CMB and Galaxy contributions ($B^\prime_\nu$ and $G_\nu$ at each channel $\nu$) one can model any measurement with more than three channels, $\nu$, 
decomposing the measured dipole, ${\cal D}$,  into the following terms:
\begin{equation}
{\cal D}^{\rm model}_\nu = {\cal D}_\nu^{\rm noise} + a G_\nu + b B^\prime_\nu + d {\cal T}_\nu^{\rm CIB}
\label{eq:dipole_future}
\end{equation}
The last term describes the CIB dipole with the template, $\cal{T}^{\rm CIB}$, given by eq. \ref{eq:cib_sed}. Given the templates, one can evaluate the CIB dipole after marginalizing over $a$ and $b$ and summing over all the available channels. This is achieved by minimizing $\chi^2=\sum_\nu ({\cal D}^{\rm sky}_\nu-{\cal D}^{\rm model}_\nu)^2/\sigma_\nu^2$ with respect to $(a,b,d)$. The solution for $d$ and its uncertainty, $\sigma_d$, is then given by standard regression and error propagation and the signal-to-noise of the prospective measurement will be given by $S/N=d/\sigma_d$. In evaluating it below we adopt $\sigma_\nu$ in $\chi^2$ to be given by the noise per multipole shown in Fig. \ref{cibde}. This allows prediction of the S/N of the CIB dipole measurements as follows for each of the three models of foreground galaxy removal in Fig. \ref{cibde} ($z\ga 0, 0.5, 1$ - dashed black, blue, green lines):
\begin{itemize}
\item {\bf FIRAS}. Substituting the parameters for the FIRAS data and the G30 template we obtain after summing over all the FIRAS channels S/N$\simeq 0.13$ for the CIB template of eq. \ref{eq:cib_sed}. This is what we have actually measured. Equation (5) shows that in order make a significant mea- surement with a future FIRAS-like instrument with the same frequency coverage, the instrument noise must be an order of magnitude below the FIRAS data noise.
\item {\bf Planck/Herchel}. These would not lead to good measurements
given that one needs to resolve three parameters ($a,b,d$) from the 2-3 (wide) frequency bands. This can already be seen from the levels of CMB residual dipole in Fig. \ref{cibde}.
\item {\bf Pixie}. Substituting the {\bf Pixie} parameters, using the G30 model template, we get 
S/N$\simeq 34, 49, 46$ for the CIB dipoles given by the dashed black, blue, green 
lines in Fig. \ref{cibde}. The S/N increases if one eliminates galaxies up to $z\simeq 0.5$ and then starts dropping again because the level of the CIB decreases too. If one uses only the parts of the
 template with the lines - at wavelengths shorter than $\simeq 160~\mu$m, starting with the C{\sc ii} line - this part of the template contributes about half the signal-to-noise when added in quadrature with S/N$\simeq 22$ for the black line of the CIB dipole (no galaxies removed). If we use the G10
  template here too we recover good prospects for such a measurement with S/N$\simeq 25, 46, 42$ for the three cases of the CIB dipole model. This argues for good prospects of this measurement with {\bf Pixie} even if not many foreground galaxies are removed from the data.
\end{itemize}

Importantly, the high S/N for a prospective CIB dipole measurement with {\bf Pixie} would allow us to also measure the dipole direction with good accuracy. The accuracy of the measured direction for $S/N\gg 1$ would be $\Delta \theta \simeq \sqrt{2} (S/N)^{-1}$radian. Thus for {\bf Pixie} the accuracy of the CIB dipole direction would be
\begin{equation}
\Delta\theta_{\rm Pixie} \simeq 2^\circ \; \frac{40}{(S/N)_{\rm Pixie}}
\label{pixie_dir}
\end{equation}
The current discrepancy between the local acceleration vector direction measured from galaxy surveys and the direction of the CMB dipole is about $\sim 15^\circ-20^\circ$ (Erdogdu et al 2006, see also Gunn 1988 for discussion of prior measurements) presenting a challenge for the purely kinematic interpretation of the CMB dipole. Thus the {\bf Pixie} measurement can settle the meaning of that discrepancy as the CIB is expected to be aligned with the true direction of the local motion.

\section{Conclusions}

We have shown that, because of its spectral energy distribution,
the far-IR CIB can provide information on the convergence and the
origin of the CMB dipole. To test this the analysis has been
applied to the best available current datasets: \COBE FIRAS and
DIRBE. The main limitation with FIRAS data is
accurate modeling of Galactic contamination because low-$z$
galaxies are similar in far-IR emission to the Galactic dust and
cannot be removed from the maps because of insufficient angular
resolution. Although the Galactic foreground has prevented us from
making a positive detection of the CIB dipole, the limit already
approaches the expected value.  Either a modest improvement in
templates (such as a better C{\sc ii} map) or in measurement accuracy
could allow a more definitive measurement.  The
{\bf Planck}/{\bf Herschel} data and, particularly, the proposed {\bf Pixie} mission 
data could provide the additional leverage to uncover the CIB dipole and probe its
alignment with that of the CMB.

\acknowledgements This work was supported by NASA 06-ADP06-0012 and NSF AST 04-06587
grants. We thank Al Kogut for providing data on the projected Pixie noise levels and the anonymous referee for useful critique of the original manuscript.


\clearpage
\begin{references}

\reference{} Atrio-Barandela, F. et al 2010, \apj, 719, 77

 \reference{} Hinshaw~G \etal 2009, ApJS 180, 225

\reference{} Bennett~C \etal 1996, ApJL, 464, L1

\reference{} Boldt~E 1987, Phys Rep, 146, 215

\reference{} Boughn, S., Crittenden, R.G. \& Koehrsen, G. P. 2002,
ApJ, 580, 672

\reference{} Devlin, M.J. et al 2009, Nature, 458, 737

\reference{} Diehl~R \etal, 1995, A\&A, 298, 445

\reference{} Erdogdu, P. et al 2006, MNRAS, 368, 1515

\reference{} Fixsen~DJ  \etal, 1994a, \apj, 420, 445

\reference{} Fixsen~DJ  \etal, 1996, \apj, 473, 576

\reference{} Fixsen~DJ  \etal, 1997, \apj, 486, 623

\reference{} Fixsen~DJ  \etal, 1998, \apj, 508, 123

\reference{} Fixsen~DJ, Bennett~CL \& Mather~JC, 1999, \apj, 526,
207  


\reference{} Fixsen~DJ \etal, 2009, \apj, submitted

\reference{} Fixsen~DJ, 2009, \apj, 707, 916

\reference{} Griffin~M \etal,  2008, Proc SPIE 7010E, 4

\reference{} Grischuk~L. 1992, Phys. Rev. D 45, 4717

\reference{} Gunn, J. 1988, In ``The extragalactic distance
scale", ASP 100th Anniversary Symposium, p. 344.

\reference{} Haslam~CGT, \etal, 1981, A\&A, 100, 209

\reference{} Hauser~MG  \etal, 1998, \apj, 508, 44

\reference{} Hinshaw~G  \etal, 2009, \apjs, 180, 225

\reference{} Itoh, Y. et al 2009, arxiv:0912.1460

\reference{}Kashlinsky, A. \& Atrio-Barandela, F.  2000,
Astrophys. J., 536, L67

\reference{} Kashlinsky, A., Atrio-Barandela, F., Kocevski, D. \&
Ebeling, H. 2008, ApJ, 686, L49

\reference{} Kashlinsky, A., Atrio-Barandela, F., Kocevski, D. \&
Ebeling, H. 2009, ApJ, 691, 1479

\reference{} Kashlinsky, A., Atrio-Barandela, F., Ebeling, H.,
Edge, A., \& Kocevski, D. 2010, ApJ, 712, L81

\reference{} Kashlinsky, A., Tkachev, I. \& Frieman, J. 1994,
Phys. Rev. Letters, 73, 1582

\reference{} Kashlinsky, A. \& Odenwald, S. 2000, \apj, 528, 74

\reference{} Kashlinsky~A. 2005, Physics Reports, 409, 361

\reference{} Kelsall  \etal, 1998, \apj, 508, 25

\reference{} Kogut~A \etal, 1993, \apj, 419, 1

\reference{} Kogut~A \etal, 2009, \apj, submitted

\reference{} Kogut~A \etal, 2010, in preparation

\reference{} Kosowsky, A. \& Kahniashvili, T. 2010, Phys Rev Lett.,
submitted (arxiv:1010.4543)

\reference{} Peebles, J. E., \& Wilkinson, D. T. 1968, Phys. Rev., 174, 2168

\reference{} Priat,~M \etal, 2002, A\&A, 393, 359

\reference{} Puget,~JL \etal, 1996, A\&A, 308, L5

\reference{} Scharf, C. et al 2000, ApJ, 544, 49

\reference{} Stark, A. et al 1992, ApJS, 79, 77

\reference{} Schlegel, D.J., Finkbeiner, D.P., Davis, M. 1998, ApJ, 500, 525

\reference{}Strauss~MA \& Willick~JA, 1995, Physics Reports,
261,271

\reference{} Tauber, J. 2004, Adv. Space Research, 34, 491

\reference{} Turner~M. 1991, Phys Rev D, 44, 3737

\reference{}Turner~M. 1992, General Relativity and Gravitation,
24, 1

\reference{} Zhang, P. 2010, MNRAS (Letters), 407, L36.
arXiv:1004.0990

\end{references}
\end{document}